\documentclass[preprint]{aastex}

\newcommand{\unit}[1]{\ifmmode\,{\rm #1}\else$\,{\rm #1}$\fi}

\begin{document}

\title{A Survey for Transient Astronomical Radio Emission at 611 MHz}

\author{C.A. Katz\altaffilmark{1,2}}
\affil{Harvard-Smithsonian Center for Astrophysics}
\affil{60 Garden St., MS 78, Cambridge, MA 02138}
\email{ckatz@cfa.harvard.edu}

\author{J.N. Hewitt\altaffilmark{1}}
\affil{Department of Physics and Center for Space Research,
Massachusetts Institute of Technology, Room 37-241,
Cambridge, MA 02139}
\email{jhewitt@mit.edu}

\author{B.E. Corey}
\affil{MIT Haystack Observatory, Route 40, Westford, MA 01886}
\email{bcorey@haystack.mit.edu}

\and

\author{C.B. Moore\altaffilmark{1,2}}
\affil{TransForm Pharmaceuticals, 29 Hartwell Ave., Lexington, MA 02421}
\email{cmoore@alum.mit.edu}

\altaffiltext{1}{formerly of Research Laboratory of Electronics, MIT,
Cambridge, MA} 
\altaffiltext{2}{formerly of Department of Physics, MIT,
Cambridge, MA}

\slugcomment{to appear in PASP}

%%%%%%%%%%%%%%%%%%%%%%%%%%%%%%%%%%%%%%%%%%%%%%%%%%%%%%%%%%%%%%%%%%%%%%%%%%%
\begin{abstract}

We have constructed and operated the Survey for Transient Astronomical
Radio Emission (STARE) to detect transient astronomical radio emission
at 611\unit{MHz} originating from the sky over the northeastern United
States.  The system is sensitive to transient events on timescales of
0.125\unit{s} to a few minutes, with a typical zenith flux density
detection threshold of approximately 27\unit{kJy}.  During 18~months of
round-the-clock observing with three geographically separated
instruments, we detected a total of 4,318,486 radio bursts.  99.9\% of
these events were rejected as locally generated interference, determined
by requiring the simultaneous observation of an event at all three sites
for it to be identified as having an astronomical origin.  The remaining
3,898 events have been found to be associated with 99 solar radio
bursts.  These results demonstrate the remarkably effective RFI
rejection achieved by a coincidence technique using precision timing
(such as GPS clocks) at geographically separated sites.  The
non-detection of extra-solar bursting or flaring radio sources has
improved the flux density sensitivity and timescale sensitivity limits
set by several similar experiments in the 1970s.  We discuss the
consequences of these limits for the immediate solar neighborhood and
the discovery of previously unknown classes of sources.  We also discuss
other possible uses for the large collection of 611\unit{MHz} monitoring
data assembled by STARE\@.

\end{abstract}

\keywords{surveys --- instrumentation: miscellaneous --- Sun:radio
radiation --- radio continuum}  

%%%%%%%%%%%%%%%%%%%%%%%%%%%%%%%%%%%%%%%%%%%%%%%%%%%%%%%%%%%%%%%%%%%%%%%%%%%
\section{Introduction}

Transient astronomical electromagnetic radiation is the signature of
some of the most fascinating physical phenomena in the universe.
Anywhere the physical conditions change with time, there is the
potential for transient radiation.  The detection of transient
astronomical radiation presents challenges not encountered in
observations of persistent sources.  Sources which produce radiation
sporadically cannot easily be studied with typical observatories and
their schedules of observing time allocation; other techniques are
required.  Detecting transient astronomical signals at radio wavelengths
in particular requires overcoming an even greater difficulty: the
pervasive presence of radio frequency interference (RFI).  The ever
growing use of wireless services means the radio spectrum is crowded
with a wide variety of signals, a large fraction of which are transient.
Since it is the very nature of transient signals of terrestrial or
astronomical origin to disappear unpredictably, they cannot reliably be
distinguished from each other by repeated observing.  Surveyors for
transient astronomical radio signals must devise other methods to
separate the desired astronomical signals from the seemingly ubiquitous
RFI\@.

We describe here the Survey for Transient Astronomical Radio Emission
(STARE), a system for detecting transient astronomical radio signals.
Sensitive to transient radio signals at 611\unit{MHz} on time scales of
0.125\unit{s} to a few minutes, STARE rejects RFI by requiring
simultaneous observation of signals at three geographically separated
sites.

\subsection{Sources of Transient Astronomical Radio Emission
\label{subsection:sources}}

Many sources have been observed to produce transient astronomical radio
emission.  Perhaps the most familiar is the Sun, from which a large
variety of transient radio signals emanate.  From microwave spike bursts
lasting $\sim10\unit{ms}$ to Type~III storms lasting weeks, the
characteristics of solar radio bursts span wide ranges of brightness
temperature, duration, frequency, and polarization.  Associated with
many different aspects of solar activity, the mechanisms producing the
bursts run the gamut, from thermal bremsstrahlung to plasma radiation.
\citet{Dulk85} and \citet{Hjellming88} provide reviews.  Other stars
have been seen to produce radio bursts as well.  Events observed on
flare stars are similar in character to those seen on the Sun, but imply
radio luminosities $10^4$ times that of solar flares \citep{Dulk85}.
RS~CVn systems, close binaries with orbital periods of
$\sim\!1$--30\unit{days}, have been observed to produce transient radio
emission on time scales of minutes to days.  The signals are thought to
be due to two separate phenomena: the acceleration of electrons in the
magnetic fields between the stars, and coherent emission from the
individual stars \citep{Hjellming88}.  X-ray binaries as well are known
to produce transient radio signals following X-ray events, although the
mechanism in this case is thought to be quite different from that of the
RS~CVn binaries \citep{Hjellming95a}.  Jupiter was discovered many years
ago to produce transient radio emission at decameter wavelengths
\citep{Burke55}, thought to be due to synchrotron radiation from
high-energy electrons trapped in the magnetic field of the planet.
Details are reviewed in \citet{Carr83}.  Brown dwarfs have been seen to
flare in the radio: a recent VLA observation of a such a flare measured
a flux density much higher than that expected from an empirical relation
between the luminosities of brown dwarf radio flares and X-ray flares
\citep{berger2001}.  Some radio pulsars are observed occasionally to
produce ``giant pulses.''  For example, about 0.3\% of the pulses from
the Crab pulsar have amplitudes greater than 1000 times the average
pulse height \citep{Lundgren95}.  High-energy cosmic rays can cause
transient radio signals at the surface of the earth.  The interaction of
high-energy particles from space and the Earth's atmosphere produces an
``extensive air shower.''  Pair production in the shower creates
populations of electrons and positrons which are systematically
separated by the magnetic field of the Earth, setting up a current which
produces a radio pulse \citep{Kahn66}.  These are only a few examples of
known sources of transient astronomical radio emission.

Other sources have been postulated, but not observed, to produce short
radio pulses.  For example, \citet{Colgate72} and \citet{Colgate75} have
predicted that a Type~I supernova should radiate an electromagnetic
pulse at radio frequencies: during the collapse, the expanding envelope
of the white dwarf acts as a ``conducting piston,'' compressing the
transverse magnetic field, thus producing a short pulse of radio
emission.  Efforts to detect such pulses are described by
\citet{Meikle78} and \citet{Phinney79}.  Another example is
``exploding'' black holes, predicted by \citet{Hawking74}: the
decrease in black hole mass due to quantum radiation causes an increase
in surface gravity, which in turn increases the emission rate.  Black
holes near the ends of of their lives would radiate intensely, releasing
$10^{30}\unit{erg}$ in the last 0.1\unit{s}.  \citet{Rees77} has
speculated that the expanding sphere of electrons and positrons would
act like the conducting piston described by Colgate for supernovas,
similarly producing a short radio pulse.  \citet{Meikle77} and
\citet{Phinney79} have reported unsuccessful searches
for these pulses.

Of particular interest for radio transient searches is the association
between high-energy emission and radio emission.  \citet{Mattox94}
reports searching the CGRO/EGRET phase~1 full-sky survey for gamma-ray
emission from X-ray selected BL~Lac objects and from the 200 brightest
radio-quiet quasars.  None was detected, while EGRET did detect
$\sim\!40$ radio-loud quasars and radio-selected BL~Lacs.  This seems to
suggest that ``apparent gamma-ray emission is intimately linked to
apparent radio-emission,'' \citep{Mattox94} which is perhaps not
surprising since the conditions which produce high-energy emission
(relativistic particles), in the presence of even a weak magnetic field,
produce radio emission through the synchrotron mechanism.  This
association has been noticed by others as well
\citep[e.g.][]{Paczynski93}.  The presence or absence of radio emission
from high-energy sources can yield clues about their workings.

Gamma-ray bursts (GRBs) have been observed to produce emission at longer
wavelengths following the gamma-ray event (see \citealp{vanParadijs2000}
for a review).  These so-called afterglows were first detected at radio
wavelengths in the GRB of 1997~May~08 \citep{Frail97c}, and the field
has since matured such that a catalog of radio afterglows is possible
\citep{Frail2003}.  The afterglows are believed to be due to emission
from shocks produced when a relativistic fireball interacts with an
ambient medium (\citealp{Meszaros2002} reviews models).  Another
possibility, which has not been detected, is a prompt radio burst
associated with the GRB event itself.  Again, energetic charged
particles in a magnetic field might serve as a source; for example,
\citet{Usov2000} suggest that strong low-frequency radio emission might
be generated by time variability in the current sheath surrounding a
magnetized jet.  Examples of other ideas relevant to the detection of
prompt radio emission from GRBs may be found in \citet{Palmer93},
\citet{Hansen2001}, and \citet{Sagiv2002}.

Energetic particles can produce radio emission under other conditions as
well.  It was suggested some time ago
\citep{Askar'yan1962,Askar'yan1965} that high energy neutrinos and
cosmic rays would generate coherent Cerenkov emission as they travel
through a dense dielectric medium such as water, ice, the earth, or the
moon.  The physical process is similar to that of the extensive air
shower discussed earlier, involving the production of a shower with an
imbalance of charge.  The spectrum of Cerenkov light from such an event
would be very broad, including radio and optical emission.  The
generation of coherent radio bursts has been verified in accelerator
experiments \citep{Saltzberg2001}.  The pulses generated by particles
traveling through the lunar regolith are expected to be very bright and
very short: more than thousands of Janskies for the highest energy
particles, with durations on the order of a nanosecond
\citep{Alvarez-Muniz2001}.  \citet{Hankins1996} report an unsuccessful
search for such emission.

\subsection{Other Work}

Experiments to detect transient astronomical radiation at radio
wavelengths have traditionally followed one of two approaches.  When
transient radio emission is thought to originate from a particular
source or region of the sky, a high-gain, small solid-angle approach is
used.  A high-gain radio telescope pointed at the region of interest
provides good sensitivity and rejection of signals outside the region.
When the location of the source of radiation is unknown, a low-gain,
large solid-angle approach is more appropriate.  Radio telescopes with
large beams provide coverage of large fractions of the sky, but at the
cost of reduced sensitivity, since the angular extent of any discrete
source is likely to be much smaller than the telescope beam.

Some good examples of the high-gain, small solid-angle approach are
those which were prompted in the early 1970s by Weber's reports of
detections of pulses of gravitational radiation
\citep[e.g.][]{Weber70b}.  Since the gravitational waves were reported
to have originated from the Galactic center, attempts to detect radio
frequency activity focused on that region.  \citet{Partridge72}
monitored the Galactic center with two radiometers separated by
100\unit{km}.  The first, at 16\unit{GHz}, provided a beamwidth of
$\sim\!12'$, sensitivity $\sim\!100\unit{Jy}$, and response time
$0.5\unit{s}$.  The other, at $19\unit{GHz}$, provided beamwidth
$\sim\!12\deg$, sensitivity $\sim\!10^6\unit{Jy}$, and response time
$3\unit{s}$.  Astronomical events were to be identified by their
simultaneous appearance in the records at both sites.  After 90~hours of
monitoring, they detected no coincidences.  \citet{Hughes73} observed
the Galactic center at 858\unit{MHz} with beamwidth $1.4\deg$,
sensitivity 85\unit{Jy}, and response time $1\unit{s}$.  In 207~hours of
monitoring the Galactic center, they report 97 detections of pulses.
However, their interference rejection scheme is unclear, making
uncertain their conclusion of the existence of discrete radio pulses
from the Galactic center.  \citet{OMongain74} used the Mount Hopkins
Observatory 10-meter optical reflector (which was fitted with two
outboard 4.6-meter radio reflectors) to make a more general search
including the Galactic center, the Coma cluster of galaxies, and the
Andromeda galaxy at three radio frequencies and one optical frequency.
In approximately 200~hours of observing, they detected no events of
extraterrestrial origin, which were to be identified by the differences
in pulse arrival times among frequencies, due to dispersion by the
interstellar medium.

Other work has attempted to observe a much larger fraction of the sky
with lower sensitivity.  For example, \citet{Charman70} assembled a
system of five receiving stations in Great Britain and Ireland with
station separations ranging from 110\unit{km} to 500\unit{km}.  Each
station had similar receiving systems at 151\unit{MHz}, consisting of
two half-wave dipole antennas operating as phase-switched
interferometers, and receivers with sensitivities of
$\sim\!10^5\unit{Jy}$ and response times of $\sim 1\unit{s}$.
Interference rejection was accomplished by requirement of five-fold
coincidence.  After $\sim\!2400$~hours of observations, they detected no
events.  \citet{Mandolesi77} used four receiving systems at Medicina
(Bologna, Italy) operating at at 151, 323.5, 330.5 and 408\unit{MHz}.
On 16~August~1976 all four instruments detected a radio burst, while
sixty seconds earlier, a gamma-ray burst was detected by three
satellites and one balloon-borne detector.  Strong rejection of local
interference was provided by an independent observation made at
237\unit{MHz} at the Astronomical Observatory of Trieste (400\unit{km}
distant from Medicina).  The investigators estimated the probability of
a random coincidence between the five-fold radio event and the gamma-ray
burst at $8 \times 10^{-5}$.  Using geometric arguments, they localized
the radio burst to a region on the sky.  Unfortunately, later work on
the gamma-ray burst data from the balloon-borne detector
\citep{Sommer78} produced a position for the gamma-ray event which was
well outside the radio emission error box, apparently ruling out a
common origin for the gamma-ray and radio events.  But the possibility
remains that the radio event was astronomical in
origin. \citet{Hugenin74} used two receiving systems separated by
several hundred kilometers to monitor the sky at 270\unit{MHz}.  Each
station consisted of a helical antenna with beam area $\sim\!1\unit{sr}$
centered on the north celestial pole, and a multi-channel receiver which
provided a sensitivity of $\sim\!10^4\unit{Jy}$ ($1\sigma$).  The data
were displayed with response times $\tau = 20\unit{ms}$ to $1\unit{s}$
on an oscilloscope, and recorded by continuously photographing the
oscilloscope screen.  Interference was rejected by requiring coincidence
between sites, and by examining for the frequency dispersion expected in
extraterrestrial signals due to the interstellar plasma.  In 213~hours
of observing, they detected no events.  \citet{Amy89} constructed the
Molonglo Observatory Transient Event Recorder (MOTER), which operated at
843\unit{MHz} in parallel with normal Molonglo Observatory Synthesis
Telescope (MOST) synthesis observations.  This system used 32
total-power fan beams spaced at full beamwidth intervals of $44''$.
When a transient event with flux density \mbox{$\gtrsim 10\unit{mJy}$}
and duration $1\unit{\mu s}$ to $800\unit{ms}$ occurred in the field of
view, MOTER compared the signals in the fan beams.  Sources closer than
about 3000\unit{km} were significantly out of focus, and thus appeared
in many beams simultaneously, while signals appearing in one beam only
were thought to be due to random noise.  In this way MOTER was able to
reject signals of local origin.  A signal which was not rejected was
localized to an arc segment on the sky corresponding to the fan beam in
which its maximum appeared.  Over a full synthesis observation, the
resulting ensemble of arc segments intersected at a point, the location
of the source.  While MOTER lacked instantaneous large solid-angle
coverage, in time much of the southern sky was surveyed.  In
$\sim\!4000\unit{hours}$ of observations, Amy, Large, \& Vaughan
detected only previously known pulsars.

\subsection{STARE}

We describe here a large solid-angle, low-gain system for detecting
transient astronomical radio emission at 611\unit{MHz} on time scales of
a few minutes or less.  Similar in spirit to the work of
\citet{Charman70} and \citet{Mandolesi77}, STARE updates previous
efforts through the use of modern technology.  The wide availability of
fast computers and other hardware permits the collection of data in
digital form.  Such a data record makes possible a variety of analyses
both during data collection and afterward, presenting a greater
opportunity for discovery than the data sets from the 1970s which were
generally in the form of analog chart records.

%%%%%%%%%%%%%%%%%%%%%%%%%%%%%%%%%%%%%%%%%%%%%%%%%%%%%%%%%%%%%%%%%%%%%%%%%%%
\section{Methods\label{section:methods}}

STARE was designed to be simple and inexpensive, and was intended to be
a first look which might eventually foster further work with new
instrumentation.  The system operates at a frequency of 611\unit{MHz} in
a bandwidth of 4\unit{MHz}, corresponding to a frequency band protected
in the United States for radio astronomy; if not protected, this band
would contain the signal for channel~37 in the UHF television spectrum.
Scientific arguments fail to indicate a clear choice of observing
frequency: at higher frequencies, optically thick sources are brighter,
while at lower frequencies, optically thin nonthermal processes radiate
more intensely.  Too low, however, and the sky brightness temperature
becomes prohibitive.  The choice of 611\unit{MHz} was a compromise among
these considerations, bolstered by technical factors such as the
relative ease and low cost of constructing radio astronomy apparatus for
this frequency range, and the existence of a protected band for
astronomy.

\subsection{Apparatus\label{subsection:apparatus}}

\begin{figure}[t]
\epsscale{0.7}
\plotone{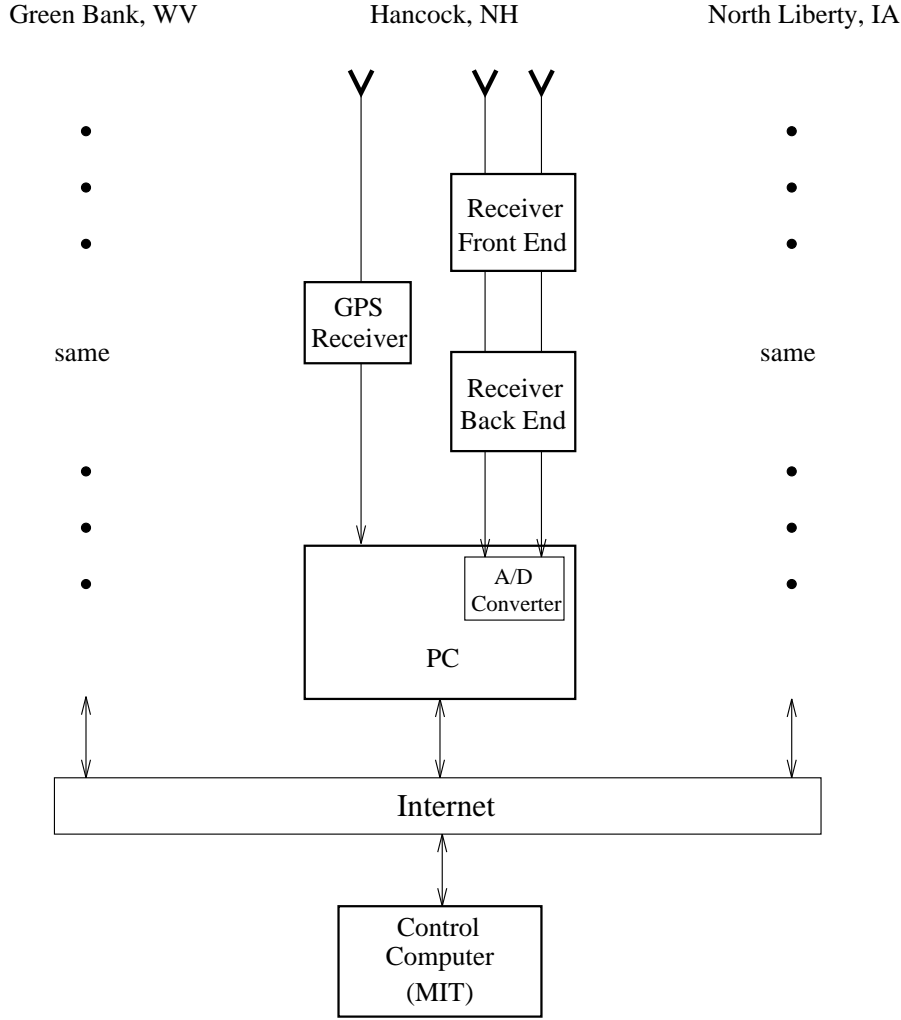}
\caption{STARE system organization. \label{fig:systemorg}}
\end{figure}

The STARE system consists of detectors at three geographically separated
sites: one at the VLBA\footnote{The VLBA is part of the National Radio
Astronomy Observatory, which is operated by Associated Universities,
Inc., under cooperative agreement with the National Science Foundation.}
station in Hancock, New~Hampshire, another at the VLBA station in
North~Liberty, Iowa, and the third at NRAO in Green~Bank, West~Virginia.
The sites are close enough that they see the same part of the sky, while
they are distant enough so that radio frequency interference at one of
the sites (from terrestrial or low-altitude transmitters) will not be
detected at the others.  This provides a powerful filter for selecting
signals of celestial origin: any signal which does not appear
simultaneously at all three sites is rejected.  These locations were
chosen because they all have operating radio telescopes and so are
expected to have low levels of RFI, and because NRAO was kindly willing
to support STARE operations.  The STARE instruments are highly
self-sufficient, requiring only a standard AC electrical power
connection, an internet connection, space for the antennas and
electronics, and a staff member willing to perform occasional minor
crisis intervention.  The apparatus at a site consists of several
systems: an analog receiving system, a Global Positioning System
receiver for providing accurate time, and a PC which collects data and
controls everything at the site.  The entire system is overseen by a
workstation at MIT, which is also where the data are archived.
Figure~\ref{fig:systemorg} shows the organization of the entire system.

\begin{figure}[t]
\epsscale{0.7}
\plotone{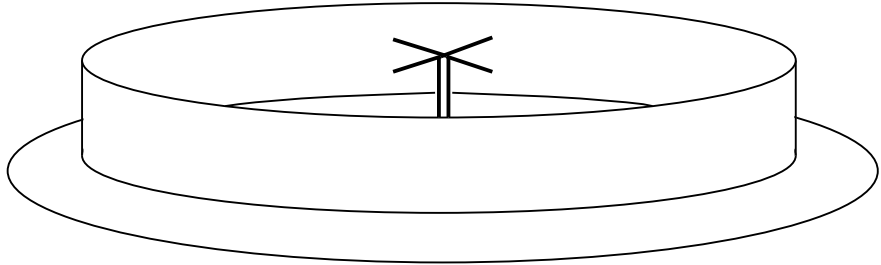}
\caption{Schematic illustration of ``crossed-dipoles in a cavity'' antenna.\label{fig:antennapicture}}
\end{figure}

The analog receiving systems each consist of an antenna, a receiver, and
a detector.  The antennas are of the ``crossed-dipoles in a cavity''
type, shown schematically in Figure~\ref{fig:antennapicture}.  This
style of antenna provides broad sky coverage, although at the expense of
sensitivity.  Laboratory measurements (by J.~Barrett, P.~McMahon, and
W.~Baumgartner) on a scale model of the antennas found a broad beam with
effective solid angle 1.4\unit{sr}, yielding a sensitivity of $6.1
\times 10^{-5}\unit{K\,Jy^{-1}}$.  The receivers are dual-channel
superheterodyne total-power radiometers, and are each divided between a
front end which is outdoors attached to the antenna, and a back end
which is indoors.  The front end begins with low-noise ambient
temperature preamplifiers \citep{Harris87} with effective noise
temperatures in the range 100-150\unit{K}\@.  The front end also
includes a laboratory-calibrated noise source for gain calibration and
system temperature measurements.  The back end performs bandpass
filtering, further amplification, frequency downconversion, square-law
detection, and anti-alias filtering at 25\unit{kHz}. The analog
receiving systems were designed to provide sensitivity to the two
orthogonal circular polarizations, with beam patterns independent of
source azimuth.  However, due to an error in the construction of the
feeds, they receive two orthogonal elliptical polarizations, which vary
with source azimuth.  We avoid this problem by summing the two received
polarizations.  It can be shown \citep{Katz97c}, and is evident from
symmetry, that the orthogonality of the received elliptical
polarizations means that their sum is independent of the azimuth of the
source.  Thus in practice we are able to calibrate only the sum of the
total power in both polarizations.

In order to reject local interference, we must be able to determine
simultaneity from site to site, requiring an accurate timekeeping
method.  This is provided by the Global Positioning System (GPS).  At
each STARE site is a GPS antenna and receiver.  When initially
installed, each GPS system was allowed to compute its position
continuously for several days.  The data were then suitably averaged to
increase the accuracy of the position determinations.  Those positions
were then programmed into the GPS systems, allowing four satellites to
be used for timekeeping alone.  In this ``static timing mode,'' the GPS
receivers have a claimed timing accuracy of $\pm 100\unit{ns}$.

Data acquisition and all other activity at each site are controlled by a
standard desktop PC\@.  The filtered square-law detector outputs are
digitized at 50\unit{kHz} by a 12-bit analog-to-digital converter card
which is clocked by a 100\unit{kHz} signal from the GPS receiver.
Timestamps are assigned to the data by switching in a short pulse from
the GPS receiver and finding the sample during which the pulse is
detected, yielding a timestamp accuracy of 20\unit{{\mu}s}.  The samples
are boxcar integrated, then collected and saved on the disk for some
specified period, then are transferred to a workstation at MIT\@.  In
addition to performing overall coordination and data archiving, the
workstation is responsible for the data analysis: the data files from
each site are analyzed individually to produce a record of transient
radio signals there.  The site event records are then compared to find
three-way coincidences.

The entire STARE system runs automatically, producing a daily report of
site events and coincidence detections.  The computer systems are set up
so that most routine maintenance can be performed without travelling to
the sites.  The system is quite robust, having required human attention
on average less than once a month.  Sometimes attention is necessary to
clear a fault condition, but most often routine maintenance
(e.g. clearing a full disk) is the cause.  Occasionally a phone call to
a site is necessary (e.g. to identify and replace an electrical fuse
blown during an electrical storm).

In normal operation, the power received by each antenna at 611\unit{MHz}
in 4\unit{MHz} bandwidth and two polarizations is boxcar averaged for
0.125\unit{s} and recorded.  Every 15\unit{min}, the noise source in
the front end is switched on for 2\unit{s}, to provide receiver gain
calibration.  Each hour, the data from the previous hour are uploaded to
the workstation at MIT, where they are processed and archived.

\subsection{Coincidence Detection\label{subsection:coincidencedetection}}

Coincidences among sites are detected by the simplest possible method:
the records of each site are examined to find events at that location,
then the lists of single-site events are compared to find instances
where all three sites recorded events simultaneously.  Note that with
this simple scheme, the overall sensitivity to three-way coincidences is
determined by the site with the poorest detection sensitivity.  If one
site fails to detect an event, the event cannot be identified as a
three-way coincidence.  More sophisticated schemes could no doubt
produce better detection sensitivity.

The first step, the detection of events at each site individually, is
performed using a two-pass sliding-window baseline fit.  On the first
pass, a quadratic model is fit to the data in a 240\unit{s} window using
a least-squares algorithm, and the dispersion $\sigma$ of the data
around this fit is calculated.  On the second pass, the model is again
fit to the data in the 240\unit{s} window, this time using the robust
estimation method described by \citet{Press92}, \S15.7.  We implemented
the method with a Lorentzian weighting distribution with the width set
to the dispersion $\sigma$ calculated in the first pass.  This method
was chosen so that we could follow the wandering baseline in the data
without our fit being skewed by outlier points, of which there are many,
including the transient signals we are trying to detect.  Note that this
results in a loss of sensitivity to events longer than a few minutes.
Greater sensitivity to long-lived events could be achieved by increasing
the width of this boxcar averaging window, at the expense of reducing
the baseline-removal effectiveness for shorter time periods.

Once the baseline fit is determined, the data are examined for samples
which deviate from the baseline by more than some threshold specified as
a multiple of $\sigma$.  When a sample is seen to deviate by more than
the threshold, an event is considered to be in progress, and the next
sample is examined.  This continues until the deviation falls below
one-half of the threshold, at which time the event is considered to have
ended.  We use this adaptive threshold to reduce the incidence of long,
temporally spiky events being broken up into multiple events as the flux
density level varies around the threshold.  Once the events are
detected, they are classified as ``single-point'' or ``multi-point,''
depending on whether the thresholds were exceeded only in a single
sample (i.e. event duration $\le 0.125\unit{s}$), or in more than one
consecutive sample.

The thresholds for event detection were chosen so that from site to
site, the flux density required to trigger an event is about the same.
Using receiver noise temperature $150\unit{K}$, zenith antenna
sensitivity $6.1 \times 10^{-5}\unit{K\,Jy^{-1}}$, integration time
$0.125\unit{\rm s}$, and bandwidth $4\unit{MHz}$, we calculate a
theoretical zenith flux density sensitivity for our systems of about
4\unit{kJy}.  Of course in practice the system temperatures are higher
than the receiver temperatures because our $1.4\unit{sr}$ beam admits
RFI from sources far and wide, degrading the system sensitivity.  We
found that the system temperatures also vary widely with time.  At
Green~Bank and North~Liberty, the median system temperatures through
18~months of observations were similar at about $200\unit{K}$.  At
Hancock, the system temperatures were typically higher by a factor of 2.
To make the zenith flux density sensitivities consistent from site to
site, we chose a threshold of $5\sigma$ for transient detection at
Green~Bank and North~Liberty, and $2.5\sigma$ at Hancock.  These
thresholds correspond to a zenith flux density triggering threshold of
about $26\unit{kJy}$.  For sources away from zenith, the threshold is of
course higher due to the decrease in antenna gain.

\begin{figure}[t]
\plotone{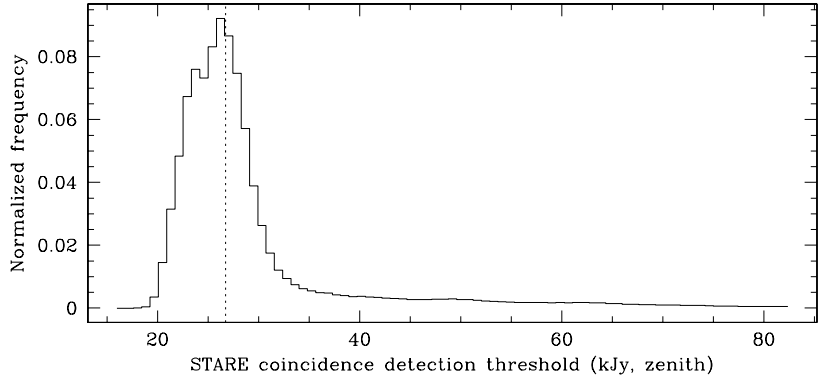}
\caption{Normalized histogram of minute-by-minute STARE zenith
coincidence detection threshold for 711,456 minutes of three-site
observations.  5.6\% of the values lie beyond the right extreme of the
plot.  The dotted line shows the median value of
26.7\unit{kJy}.\label{figure:coincidencesensitivity}}
\end{figure}

Once the single-site event lists are produced, they are compared to find
events detected simultaneously at all three sites.  Three-way
coincidences are classified as ``single-point'' if the three single-site
events are all single-point events, ``multi-point'' if they are all
multi-point events, and ``mixed-type'' if the simultaneous single-site
events are a combination of single-point and multi-point.  To assess the
coincidence detection sensitivity of STARE, we examined the flux density
sensitivity at each site for every minute of time that all three sites
were on-line.  We then chose the poorest sensitivity of the three as the
STARE coincidence sensitivity, since the site with the poorest
sensitivity would determine the overall sensitivity.  The results are
shown in a normalized histogram in
Figure~\ref{figure:coincidencesensitivity}.  Not shown are the 5.6\% of
the values which lie beyond the right side of the plot, due to noisy
times at one or more of the sites.  The median overall STARE zenith
sensitivity of 26.7\unit{kJy} is very close to the median single-site
detection sensitivity of 26\unit{kJy}, indicating that most of the time,
the three sites run with comparable sensitivities.

\subsection{Calibration\label{subsection:calibration}}

Event data were calibrated in several steps.  First, the receiver gains
were determined using the noise source on/off data.  These were used to
convert the data to antenna temperatures in each polarization channel.
The antenna temperatures were summed to find the total antenna
temperature, and to avoid the problem with the unknown polarizations of
the feeds (see the discussion of the analog receiving systems in
\S\ref{subsection:apparatus}).  Then, using the antenna sensitivity
determined from measurements on a scale model of the antennas, the
antenna temperatures were converted to the flux densities that would be
measured if the source were at the zenith.  This resulted in a lower
limit on the flux density of the source.  For sources of known
elevation, such as the Sun, the zenith flux densities were corrected for
the elevation response of the antenna, yielding a source flux density
measurement.

The calibration is rather coarse.  In practice we find that the flux
densities of solar radio bursts measured by the STARE systems generally
agree from site to site only within a factor of 2 or so.  Comparing
measurements between sites, we find that the Green~Bank instrument
systematically reports flux densities approximately a factor of 2 larger
than those reported by Hancock, which in turn are approximately a factor
of 2 larger than those reported by North~Liberty.  In addition, there is
significant variation of these ratios from event to event.  We believe
this is due primarily to uncertainties in two parts of the receiving
systems.  First, the beam patterns of the antennas were determined on a
$1/10$ scale model, and may not transfer very well to the actual feeds.
This would cause elevation dependent variations in the calibration and
could account for the scatter in the systematic differences between
sites.  Second, the excess noise temperatures of the noise sources used
for gain calibration were measured when the systems were constructed,
and may have changed over the several years of aging they have
experienced.  This could be the cause of the overall systematic
differences between sites.

Clearly the calibration accuracy could be improved by measuring or
computing numerically the antenna beam patterns, and by measuring the
present noise source characteristics.  But for our purpose here, we
consider the calibration accuracy of a factor of a few to be adequate
and defer the improvements to later work.

%%%%%%%%%%%%%%%%%%%%%%%%%%%%%%%%%%%%%%%%%%%%%%%%%%%%%%%%%%%%%%%%%%%%%%%%%%%
\section{Results}

The first STARE site was set up at the VLBA station in Hancock, NH\@.  A
subsequent period of site evaluation and debugging led to the final
configuration with sites at Hancock, New Hampshire; North Liberty, Iowa;
and Green Bank, West Virginia.  The system operated in this state for
approximately 18 months.  We describe here the results obtained from the
data collected from 1998~May~27 through 1999~November~19.

\subsection{Single-Site Event Detection\label{subsection:singlesitedetection}}

In 18~months of operation, STARE detected hundreds of thousands of
events at Green~Bank and North~Liberty, and millions at Hancock with its
weaker triggering threshold.  The exact numbers are given in
Table~\ref{table:singlesiteresults}.

\begin{table}[t]
\begin{center}
\begin{tabular}{llll}
                        & Hancock      & Green Bank     & North Liberty     \\ \hline
detection threshold  & $2.5\sigma$ & $5\sigma$  & $5\sigma$ \\ \hline
no.\ of single-point events  & 3,654,485 & 183,851  & 100,692      \\
mean single-point event rate & $281.5\unit{{\rm hr}^{-1}}$ &
$14.2\unit{{\rm hr}^{-1}}$ & $7.8\unit{{\rm hr}^{-1}}$      \\ \hline
no.\ of multi-point events   & 168,833 &95,496   & 115,129     \\
mean multi-point event rate & $13.0\unit{{\rm hr}^{-1}}$ & $7.4\unit{{\rm
hr}^{-1}}$ & $8.9\unit{{\rm hr}^{-1}}$      \\ \hline
\end{tabular}

\caption{\label{table:singlesiteresults}Single-site event detection
results, 1998 May 27 to 1999 November 19}
\end{center}
\end{table}

Using the mean single-point event rates computed from the single-site
event detection results, we can estimate the rate of accidental
single-point coincidences among the sites.  For a mean event rate $r$,
the probability that a time interval $\delta t$ contains an event is
$r\delta t$ (assuming $r \ll 1/\delta t$, which is true for STARE since
$\delta t = 0.125\unit{s}$).  Thus for three sites with mean event
rates $r_1$, $r_2$, and $r_3$, the probability that a time interval
$\delta t$ contains an event at all three sites is $r_1 r_2 r_3 (\delta
t)^3$.  The mean time $\Delta T$ between accidental three-site
coincidences is then $\Delta T = 1/r_1 r_2 r_3 (\delta t)^2$.  Using the
rates given in Table~\ref{table:singlesiteresults}, we find $\Delta T
\approx 3\unit{\rm years}$.  With only two sites (even those with the
lowest event rates: Green~Bank and North~Liberty), this time is $\Delta
T= 1/r_1 r_2 \delta t \approx 11\unit{\rm days}$.  This illustrates the
power of the coincidence requirement in filtering out local
radio frequency interference.  It also makes clear that the third site is
required to reduce the accidental coincidence rate to a manageable
level.

A similar examination of accidental coincidence rates for multi-point
events is not necessary, since their temporal structure provides much
more information than is available with single-point coincidences.
Accidental coincidences may be ruled out simply by direct comparison of
the time-series data from each of the sites.  Unless the events have the
same shape, they can be rejected as events of interest.

\subsection{Coincidence Detection}

In approximately 18 months of data collection, the three STARE sites
recorded well over 4~million radio bursts.  Of these, 3898 were
identified to be in temporal coincidence among all three sites: 1859 at
Hancock, 1069 at Green Bank, and 970 at North Liberty.  The numbers are
unequal because of the spiky temporal nature of many of the events.
Despite the use of the adaptive threshold algorithm described in
\S\ref{subsection:coincidencedetection}, the event detection algorithm
tended to break up long events into multiple events, depending on
exactly how the measured power varied around the detection thresholds.
In many instances, a single long event observed at one site was broken
up into multiple events at the others, due to sensitivity differences
between sites.  In this case, STARE reported multiple coincidences
despite one site reporting only a single event.  To account for this
effect, the data for each reported coincidence were examined visually to
determine which of these multiple events were really just parts of the
same overall event.  After this reduction step, we found that the
reported coincidences collapsed into 126 distinct events.  Examining
each of these more carefully, we found that 27 were accidental.
Accidental events were identified using two criteria: the time
dependence of the events differed obviously among the sites, and/or one
or more of the sites showed a temporarily very high event rate, due to
some local RF emitting phenomenon.  Ignoring the accidental events, we
find that STARE detected 99 events which appear to be of astronomical
origin.

Since we expect that the Sun is the most intense source of transient
radio emission in the sky, we compared the STARE events with those
detected independently by a solar monitoring station.  The United States
Air Force operates a worldwide ``Radio Solar Telescope Network'' (RSTN)
for the purpose of producing warnings about solar weather events which
could disrupt terrestrial systems.  Conveniently for us, one of the
stations of the RSTN is Sagamore Hill Solar Radio Observatory in
Hamilton, Massachusetts, approximately 80\unit{km} distant from our
STARE instrument in Hancock, New Hampshire.  The RSTN monitors the Sun
for transient activity at eight fixed frequencies, one of which is
610\unit{MHz}, the same as the STARE observing frequency.  Thus the
record from the Sagamore Hill RSTN station is useful to us for
comparison and verification purposes.\footnote{Data from the RSTN are
available through the National Geophysical Data Center of the National
Oceanic and Atmospheric Administration.}  The RSTN 610\unit{MHz} system
operates on an 8.5\unit{m} dish, using a dipole feed to measure a single
linear polarization, recording the solar flux density once per second
\citep{Heineman97}.

\begin{figure}[t]
\plotone{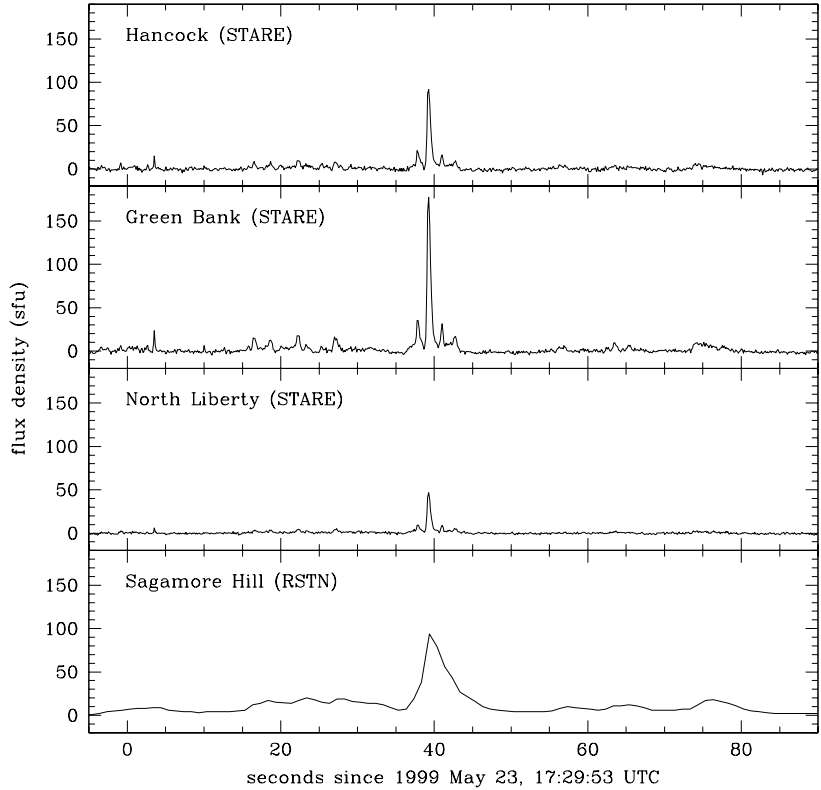}
\caption{Solar radio burst detected by STARE and RSTN at 610\unit{MHz}
on 1999 May 23.  Flux densities are in solar flux units 
($1\unit{sfu}=10^4\unit{Jy}$).\label{figure:eventdetail}}    
\end{figure}

During the 18~months of STARE three-site data collection, the RSTN
Sagamore Hill station reported 146 solar radio bursts at 610\unit{MHz}.
After combining these events with the 99 detected by STARE during this
time, we divided the ensemble of events into three groups: those
detected by STARE only (58 events), those detected by RSTN only (105
events), and those detected by both STARE and RSTN (41 events).  Of
course the first group is potentially the most interesting as it may
include transient astronomical radio emission from non-solar sources.
The other groups are useful in understanding the behavior of the STARE
system.  To illustrate a typical event, we show in
Figure~\ref{figure:eventdetail} the event detected by STARE and RSTN on
1999~May~23 at 17:30~UTC.

We first compared the peak flux densities of the events measured by
STARE and RSTN, and we found that the Hancock site produces flux density
estimates which are on average closest to those of the RSTN\@.  The
Green~Bank site produces higher values, and the North~Liberty site
produces lower values, more or less consistent with the systematic flux
density scale differences described in
\S\ref{subsection:calibration}, although with a large scatter
around these averages.  Note however that we do not expect particularly
good agreement between STARE and RSTN flux density measurements since
STARE measures the total flux density in both polarizations, while RSTN
measures a single linear polarization.  The exact ratios of the
measurements depend on the polarization of the received radiation.

\begin{figure}[t]
\plotone{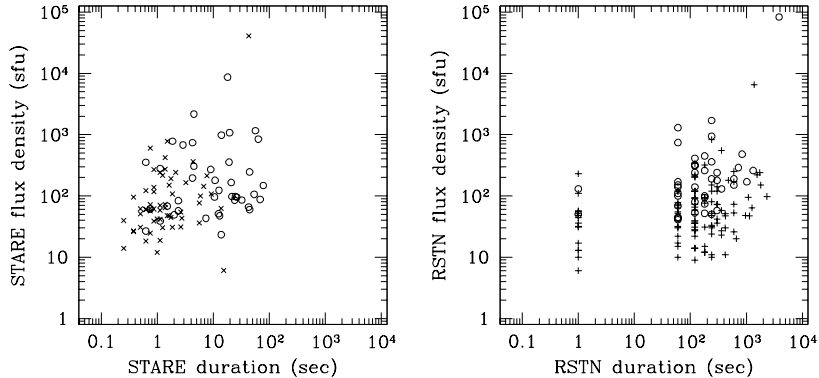}
\caption{\label{figure:durvsflux}Measured flux densities
($1\unit{sfu}=10^4\unit{Jy}$) of events vs.\ event durations, using
STARE-measured values (left) and RSTN-measured values (right).  The
circles ($\circ$) indicate events detected by both RSTN and STARE,
 the x symbols ($\times$) indicate events detected by STARE only, and the
plus symbols ($+$) indicate events detected by RSTN only.}
\end{figure}

We used the full ensemble of 204 events to help characterize the
triggering criteria of STARE and RSTN\@.  Figure~\ref{figure:durvsflux}
shows plots of event flux density against event duration for both
RSTN-measured values and STARE-measured values (from Hancock, since it
is in closest agreement with RSTN).  We can draw several conclusions
from these plots.  From the left plot, we see that STARE is more
sensitive to short duration events.  This is to be expected, since STARE
takes eight samples per second, while the RSTN takes one.  From the
right plot, we see that the RSTN has better sensitivity.  This is also
to be expected, since RSTN uses a parabolic dish which tracks the Sun.
To properly interpret the right plot in Figure~\ref{figure:durvsflux},
note that RSTN reports only the start and end times for events, with
one-minute accuracy, accounting for the distinct columns of points in
the plot.  For this plot, we arbitrarily assigned a duration of
1\unit{s} to events listed as starting and ending in the same minute.

We then examined the 105 events detected by RSTN only, and found three
reasons that they were not detected by STARE\@.  74 of the events were
simply too faint for detection by STARE\@.  For another 25 of the
events, manual examination of the STARE data showed that the events were
indeed recorded, but that they did not exceed the trigger threshold at
all three sites, so they were not identified as 3-way coincidences.  For
the remaining 6 events, there were no data recorded by one or more STARE
sites at the time of the RSTN event, due to the systems being off the
air for maintenance or other purposes.  From these results we gain
confidence that during the 18~months of three-site data collection,
STARE detected the events we expected it to detect.

Finally, we examined the final group of 58 events: those detected by
STARE but not by the RSTN\@.  From the left plot in
Figure~\ref{figure:durvsflux}, we see that many of these events are of
short duration, suggesting that they might have been of solar origin but
were too short to trigger a radio burst alert from RSTN\@.  In addition,
all of the events occurred during daytime hours.  To determine
unequivocally whether these events were from the Sun, we obtained the
RSTN 1-second data stream and compared it directly with the STARE
records.  For 51 of the events, the signal was easily discernible in the
Sagamore~Hill RSTN record, matching the STARE event well in time and
shape.  A somewhat deeper investigation into why these events were not
reported by RSTN determined that when events are detected by RSTN, a
human examines the data to decide whether burst alerts should be issued
\citep{Heineman97}, so we should not count on the burst alerts as a
complete record.  The other 7 events were not found in the Sagamore~Hill
RSTN record because of gaps in the data, due to equipment outages or
calibration.  For these, we obtained and examined the data from two
other RSTN sites: San~Vito, Italy, and Palehua, Hawaii.  Five of the
remaining events were unambiguously identified in these records.  The
two remaining events occurred at times when no RSTN data were available;
at San~Vito, the sun had already set, and at Palehua, the events
occurred during gaps in the data.  However, we suspect that these two
are due to solar radio bursts as well.  For one event, the Sagamore~Hill
data resume several seconds after the STARE event time, and show the
final moments of an event in progress.  For the other, a large solar
radio burst (detected by both STARE and RSTN) occurs less than
30~minutes after the event.  In both the RSTN and STARE records we see
that events tend to be clustered in time, so it would not be unusual if
this event were related to the following large burst.  In addition, both
of the unidentified STARE events are very short ($<0.5\unit{s}$), and
so are unlikely to have been detected by RSTN, as discussed above in
reference to Figure~\ref{figure:durvsflux}.  Although we cannot
definitively associate these two events with solar radio bursts, we
believe that the indirect evidence indicates that such an association is
warranted.  From these results we deduce that all of the astronomical
signals detected by STARE were due to solar radio bursts.

%%%%%%%%%%%%%%%%%%%%%%%%%%%%%%%%%%%%%%%%%%%%%%%%%%%%%%%%%%%%%%%%%%%%%%%%%%%
\section{Discussion}

We have presented evidence that all of the astronomical events detected
by STARE are of solar origin.  With this result we can speculate about
astrophysical scenarios that could result in STARE detections.  To do
this we establish a fiducial flux density detection threshold for the
system.  From the results in \S\ref{subsection:coincidencedetection}, we
begin with $26.7\unit{kJy}$, the observed median STARE zenith
coincidence detection sensitivity.  Then, since most events would happen
away from zenith, we multiply by 3, the approximate factor by which the
antenna response is reduced at $45^{\circ}$ elevation.  This yields
approximately $80\unit{kJy}$, which we adopt for this section as the
typical STARE coincidence detection sensitivity.

\begin{figure}[t]
\plotone{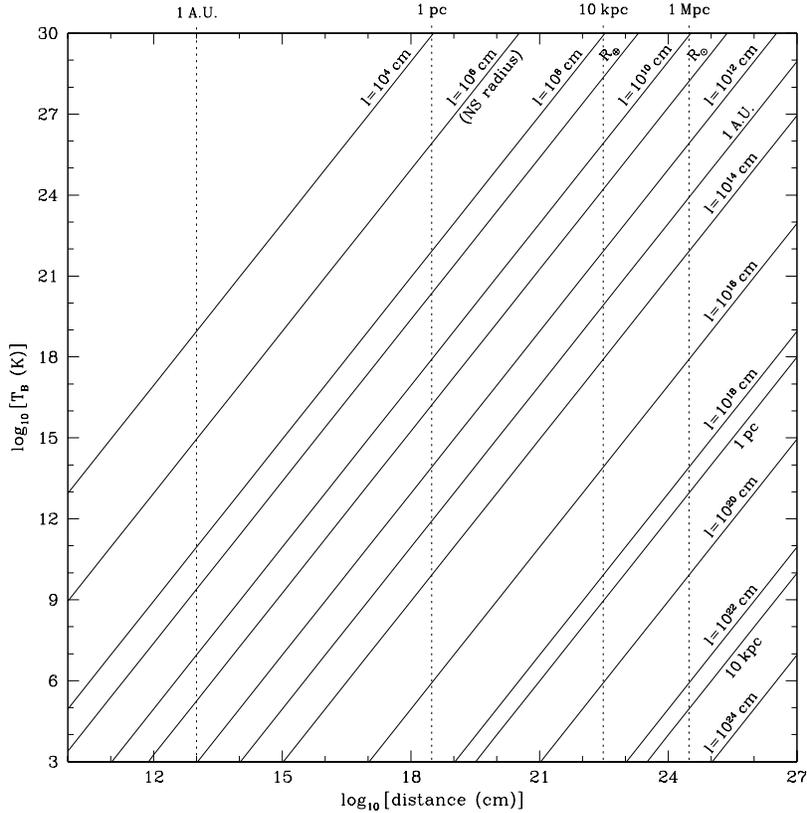}
\caption{\label{figure:brightnesstemp}Brightness temperature required of
a source to produce, under typical conditions, a detection by STARE: any
combination of source linear size (diagonal lines) and distance
(horizontal axis) yields the brightness temperature which produces a
flux density greater than $80\unit{kJy}$.}
\end{figure}

To interpret this limit in an astrophysical context, it is useful to
recast it in terms of brightness temperature $T_B$.  For a source with a
uniform spatial brightness distribution, the flux density is $S_\nu =
I_\nu\Omega$, where $I_\nu$ is the specific intensity of the source, and
$\Omega$ is its solid angle. Then in the Rayleigh-Jeans limit ($h\nu \ll
k_BT$), the source has brightness temperature
\begin{equation}
T_B = \frac{c^2 S_\nu}{2k_B\nu^2\Omega}.
\end{equation}
Using our detection limit fixes $S_\nu \sim 80\unit{kJy}$.
An object of linear size $\ell$ at distance $d$ occupies a solid angle
$\Omega \sim \frac{\pi}{4}\left(\frac{\ell}{d}\right)^2$, yielding
\begin{equation}
T_B \approx \left(\frac{2c^2S_\nu}{\pi k_B\nu^2}\right)
        \left(\frac{d}{\ell}\right)^2
      =
        8.9\unit{K} \left(\frac{d}{\ell}\right)^2
\end{equation}
Figure~\ref{figure:brightnesstemp} shows this relation plotted for
several choices of $\ell$.  For a source of linear size $\ell$ at a
distance $d$, this plot indicates the brightness temperature required to
produce a flux density which would trigger a detection by STARE\@.
It is immediately obvious that at this sensitivity STARE has no hope of
detecting distant objects.  For nearby objects, say within a few kpc,
the required brightness temperatures are high but not unprecedented.
For example, we have shown the unequivocal detection of solar radio
bursts.  Typical flux densities of these events are $\sim 1\unit{MJy}$,
corresponding to a brightness temperature of
\begin{equation}
T_B \approx \frac{2c^2S_\nu d^2}{\pi k_B\nu^2 \ell^2}
      =
        5.1\times 10^6\unit{K} \left(\frac{1R_{\sun}}{\ell}\right)^2.
\end{equation}
Many of the observed solar radio bursts display intensity variations at
the STARE time resolution of 0.125\unit{s}, which with a light
travel-time argument yields an upper limit on the size of the emission
region of $\ell \le 37,500\unit{km} \approx 0.05R_{\sun}$.  Using these
values, we find a brightness temperature limit of $T_B \gtrsim 2\times
10^{9}\unit{K}$.  This combination of duration and brightness suggests
that the detected events are Type~I or Type~III solar bursts, both of
which arise from coherent plasma radiation \citep{Dulk85}.

Although it is unlikely that any members of known classes of non-solar
sources would be near enough for detection by STARE, it is illustrative
to consider what conditions would be required of them for detectability.
One example of a class of source known to produce very high brightness
temperatures is that of the radio pulsars.  As mentioned in
\S\ref{subsection:sources}, the Crab pulsar has been observed to produce
individual giant pulses with flux densities exceeding 2000\unit{Jy}.
With observing frequency $\nu = 800\unit{MHz}$ \citep{Lundgren95},
distance $d = 2\unit{kpc}$ to the pulsar, and choosing $\ell =
100\unit{km}$ for the size of the emission region, the brightness
temperature is
\begin{equation}
T_B \ge \frac{2c^2 S_{\nu} d^2}{\pi \nu^2 k_B \ell^2} \approx 8\times
  10^{28}\unit{K}.
\end{equation}
The steep spectrum of pulsar emission would make this even brighter at
611\unit{MHz} \citep{Lyne90}.  This is close to the detection threshold
for STARE, but since the pulses are significantly shorter than the STARE
averaging time, the signal would be too diluted to detect.  However,
this suggests that another closer pulsar which produced such
bright giant pulses might be detectable.

We can consider more generally the radio emission due to a particular
emission mechanism.  For example, many sources produce radio emission at
611\unit{MHz} by the synchrotron mechanism, in which high-energy
electrons are accelerated through a spiral path by a magnetic field.  It
is well known that in general a source emitting incoherent synchrotron
radiation cannot shine with brightness temperature greater than $T_B
\sim 10^{12}\unit{K}$, since inverse Compton scattering becomes the
dominant energy sink \citep[e.g.][]{Shu91}.  From
Figure~\ref{figure:brightnesstemp} we see that a synchrotron source with
$T_B = 10^{12}\unit{K}$ would only be detectable by STARE if it were
improbably large and close.  For example, a source of size 1\unit{A.U.}
would have to be within a few tens of parsecs to produce a flux density
above the STARE detection threshold.  However, it is possible for such
sources to exceed the $T_B \sim 10^{12}\unit{K}$ limit temporarily
during a phenomenon which causes an impulsive injection of energy and a
corresponding flare \citep{Hughes91}.  One class of source which is
known to radiate via the synchrotron mechanism and produce rapid
variability is that of the Galactic ``microquasars'' \citep{Mirabel99}.
\citet{Rodriguez95} monitored the microquasar GRS~1915+105 for several
months, observing large outbursts, reaching a maximum of 1.5\unit{Jy} at
1.4\unit{GHz}.  Using this flux density and the observed spectral
power-law index of -0.87, we obtain the expected flux density at
611\unit{MHz}; with the source distance of 12.5\unit{kpc} and assuming a
source size of 1\unit{A.U.}, we can calculate a brightness temperature
of $T_B \approx 10^{15}\unit{K}$ for this outburst.
Figure~\ref{figure:brightnesstemp} shows that with these physical
parameters, such an outburst might be detectable by STARE if it were
closer than a kiloparsec or so.

To view the radio sky somewhat differently, STARE could operate with a
shorter boxcar averaging period than the 0.125\unit{s} (6250 samples of
20\unit{\mu s} each) used for the data presented here, yielding better
time resolution (at the expense of sensitivity, of course).  If the
averaging length were shortened to less than the light travel time from
one STARE station to another, the localization of sources in the sky
would be possible in principle by comparing arrival times of signals at
the STARE stations.  In addition to source positions, this method would
provide another filter for rejecting terrestrial interference.  At the
highest data acquisition rate (50\unit{kHz}), the three STARE stations
could theoretically produce localizations of transient astronomical
radio sources to better than $1^{\circ}$ over much of the sky
\citep{Katz97c}.  In practice, operating STARE in such a mode would
require further technical development to handle the high data rate
described.  With the dizzying pace of computer technology advancement,
however, the required technical improvements are much more tractable at
this writing than they were just a few years ago when STARE was
constructed.  A burst localization mode like that described here will
likely be an important feature of future work.

STARE has assembled in digital form a large record of the temporal
behavior of the radio sky at 611\unit{MHz}, presenting wide opportunity
for further analysis.  One possibility is to take advantage of the
multiple sampling of the same region of sky every 24~hours by averaging
and performing a frequency domain analysis to search for periodic
signals.  This would be very much in the same vein as the work of
\citet{Chakrabarty95} and \citet{Bildsten97}, who analyzed the archived
BATSE 1.024\unit{s} data stream to find previously unknown X-ray
pulsars.  While the sensitivity of the BATSE large-area detectors is
well below its optimum at the typical peak energies of X-ray pulsars,
the large volume of data collected by BATSE allowed excellent
sensitivity by averaging.  The situation is much the same with STARE\@.
By averaging many months worth of data, good sensitivity can in
principle be achieved for periods greater than 0.25\unit{s}.  A
significant number of radio pulsars have periods longer than this, with
many exceeding 1\unit{s} \citep{Taylor93}.  Unknown radio pulsars with
such long periods would be candidates for detection by STARE through
this technique.

%%%%%%%%%%%%%%%%%%%%%%%%%%%%%%%%%%%%%%%%%%%%%%%%%%%%%%%%%%%%%%%%%%%%%%%%%%%
\section{Conclusions}

We have operated the Survey for Transient Astronomical Radio Emission
from 1998~May~27 to 1999~November~19, monitoring the 611\unit{MHz} radio
sky to detect transient radio signals of astronomical origin with
durations of a few minutes or less.  STARE observed the sky above the
northeastern part of the United States using three geographically
separated zenith-looking detectors which made 8 measurements per second,
24 hours per day.  In 18~months of observing we detected a total of
4,318,486 radio bursts at the three STARE stations.  Of these, 99.9\%
were determined to be due to local sources of radio noise.  The
remaining 3,898 were found to be associated with 99 solar radio bursts
observed in coincidence at the three stations.  These results
demonstrate the remarkable effectiveness of an RFI discriminator based
on a coincidence technique using precision timing (such as the GPS) at
geographically separated sites.

Technological advances have allowed STARE to update similar experiments
performed in the 1970s.  With the STARE data stored digitally, the
potential for further analysis is much greater than for those
experiments which recorded data on chart paper.  The data could easily
be reanalyzed to detect transients on longer time scales.  In addition,
the large temporal and solid angle coverage could make the data useful
for a variety of other purposes, such as searches for periodic signals
and studies of the radio frequency interference environment at the
observatory sites.

We can interpret the non-detection of extrasolar transient astronomical
radio emission as indicative of the absence of bright bursting or
flaring sources within 1\unit{kpc} or so of the solar system, over the
18~months of our observations.  This rules out the existence of any
known classes of sources in the solar neighborhood.  Perhaps more
importantly, it rules out the existence of nearby sources of previously
unknown types which might produce short timescale transient radio
emission.  The absence of quiescent radio emission from such sources
means that they would not have been detected by traditional pencil-beam
surveys; through their flaring activity they may have revealed their
existence only to a program such as STARE\@.

The detection of transient astronomical electromagnetic radiation
requires different observing techniques than those offered by the
typical observatory with its scheduled blocks of time.  Since signals
may appear unexpectedly in time and space, their detection requires a
different class of temporal and spatial coverage than that provided by
traditional telescopes.  In a simple system like STARE, spatial coverage
is achieved at the expense of reduced brightness sensitivity, while
temporal coverage is achieved by the use of dedicated automated
unattended instruments which can monitor continuously.  STARE has been a
useful exercise for exploring the techniques required to detect
unexpected signals from astronomical sources.  We consider this to be
preliminary work in this area, which we expect will progress to
development of more sophisticated techniques and instruments providing
better coverage with higher sensitivity.  Indeed, we observe in the
planning for new radio telescopes such as the Square Kilometer Array and
the Low-Frequency Array for Radioastronomy renewed interest in the
transient radio sky, with discussions of sources such as neutron star
magnetospheres, gamma-ray burst sources, planetary magnetospheres and
atmospheres, accretion disk transients, and even extraterrestrial
intelligence, to name a few.  And of course the most exciting prospect
for such work is the detection of entirely new types of sources.  It
appears that the detection of transient astronomical radio emission will
be a topic of great interest for many years to come.

%%%%%%%%%%%%%%%%%%%%%%%%%%%%%%%%%%%%%%%%%%%%%%%%%%%%%%%%%%%%%%%%%%%%%%%%%%%
\acknowledgements

We thank the many people who have made STARE possible.  At the National
Radio Astronomy Observatory, M.~Goss offered the use of VLBA stations
for STARE receivers, while F.J.~Lockman offered the use of space at
Green~Bank.  Development and operation of STARE would have been
impossible without the help of T.~Baldwin and D.~Whiton at the Hancock
VLBA station, D.J.~Beard and B.~Geiger at the North~Liberty VLBA
station, and M.~McKinnon and R.~Creager at Green~Bank. Some initial
STARE testing was performed at Hat~Creek Radio Observatory, where space
and support were offered by D.~Backer, and the work was aided by
R.~Forster and M.~Warnock.  Development of the STARE instrumentation was
helped along by advice from A.~Rogers, L.~Beno, and P.~Napier.
J.~Ellithorpe contributed to the computer work, and J.~Barrett provided
many hints and lessons.  H.~Coffey and E.~Irwin at the National
Geophysical Data Center provided the data from the RSTN telescopes, and
TSgt~C.~Hoffman, U.S.A.F., of the Sagamore~Hill RSTN station answered
many questions about RSTN operations and methods.  Finally, we thank the
anonymous referee whose insightful comments led to significant
improvements in this article.

We are grateful for the support STARE has received from a variety of
sources, including a David and Lucille Packard Fellowship in Science and
Engineering, a Bruno~Rossi Graduate Fellowship in Astrophysics, the MIT
Class of 1948, a National Science Foundation Presidential Young
Investigator grant (AST-9158076), and a Postdoctoral Fellowship from the
Smithsonian Institution.

%%%%%%%%%%%%%%%%%%%%%%%%%%%%%%%%%%%%%%%%%%%%%%%%%%%%%%%%%%%%%%%%%%%%%%%%%%%
% references

\end{document}